\documentclass[12pt]{iopart}

\usepackage{etoolbox}
\usepackage{graphicx}
\usepackage{epstopdf}
\usepackage{amssymb}
\usepackage{iopams}
\usepackage{color}
\usepackage{xspace}%to get spacing right in new commands like e.g.
\newcommand\etc{\textit{etc.}\xspace}
\newcommand\eg{\textit{e.g.}\xspace}
\newcommand\ie{\textit{i.e.}\xspace}

\renewcommand\Im{\mbox{$\mathrm{Im}$}}

\def \Bo {B}

\def \kpar {\mbox{$k_{\parallel}$}}
\def \kperp {\mbox{$k_{\perp}$}}

\def \Dkx {\mbox{$\Delta k_{x}$}}
\def \Dky {\mbox{$\Delta k_{y}$}}
\def \Dkz {\mbox{$\Delta k_{\parallel}$}}
\def \bkperp {\mbox{${\bf k}_{\perp}$}}
\def \Lpar {\mbox{$L_{\parallel}$}}
\def \vT {\mbox{$v_{\scriptsize{\mathrm{T}}}$}}

\newcommand{\vpar}{\ensuremath{v_{\parallel}}}
\newcommand{\vperp}{\ensuremath{v_{\perp}}}

\def \kyo {\mbox{$k_{y}^{\mathrm{o}}$}}
\def \ello {\ell^{\mathrm{o}}}

\def \phiHz {\hat{\phi}^{\mathrm{z}}}
\def \THz {\hat{T}^{\mathrm{z}}}

\def \os {\mbox{$\omega_*$}}
\def \osT {\mbox{$\omega_*^{\mathrm{\scriptscriptstyle T}}$}}
\def \ost {\mbox{$\tilde{\omega}_*$}}

\def \od {\mbox{$\omega_d$}}
\def \odt {\mbox{$\tilde{\omega}_d$}}
\def \omegaNL {\mbox{$\omega_{\mathrm{NL}}$}}
\def \omegaNLz {\mbox{$\overline{\omega}_{\mathrm{NL}}$}}

\def \gammaL {\mbox{$\gamma_{\mathrm{L}}$}}

\def \oG {\mbox{$\omega_\mathrm{G}$}}

\newcommand{\zon}[1]{\overline{#1}}
\newcommand{\nzo}[1]{\tilde{#1}}

\def \Ephi {\mbox{$E_{\phi}$}}
\def \nEphi {\mbox{$\nzo{E}_{\phi}$}}
\def \zEphi {\mbox{$\zon{E}_{\phi}$}}

\newcommand{\citep}[1]{\cite{#1}}
\newcommand{\citet}[1]{\cite{#1}}

\begin{document}

\title[Understanding saturation in ion temperature gradient turbulence]{Understanding nonlinear saturation in zonal-flow-dominated ion temperature gradient turbulence}
\author{G. G. Plunk}
\ead{gplunk@ipp.mpg.de}
\address{Max Planck Institute for Plasma Physics, Wendelsteinstra{\ss}e 1, 17491}
\author{A. Ba\~{n}\'{o}n Navarro}
\address{Max Planck Institute for Plasma Physics, Boltzmannstra{\ss}ee 2, 85748 Garching, Germany}
\address{Department of Physics and Astronomy, University of California, Los Angeles, California 90095}
\author{F. Jenko}
\address{Max Planck Institute for Plasma Physics, Boltzmannstra{\ss}e 2, 85748 Garching, Germany}
\address{Department of Physics and Astronomy, University of California, Los Angeles, California 90095}

\begin{abstract}
We propose a quantitative model of ion temperature gradient driven turbulence in toroidal magnetized plasmas.  In this model, the turbulence is regulated by zonal flows, \ie mode saturation occurs by a zonal-flow-mediated energy cascade (``shearing''), and zonal flow amplitude is controlled by nonlinear decay.  Our model is tested in detail against numerical simulations to confirm that both its assumptions and predictions are satisfied.  Key results include (1) a sensitivity of the nonlinear zonal flow response to the energy content of the linear instability, (2) a persistence of zonal-flow-regulated saturation at high temperature gradients, (3) a physical explanation of the nonlinear saturation process in terms of secondary and tertiary instabilities, and (4) dependence of heat flux in terms of dimensionless parameters.
\end{abstract}

\maketitle

\section{Introduction}

Ion temperature gradient (ITG) driven turbulence is typically one of the largest causes of thermal losses in magnetized fusion plasmas.  It is also an intensely studied type of plasma turbulence, for which a quantitative physical understanding is within grasp.  Decades of contentious research seem to converge on a single fact: the interaction between plasma waves and spontaneously generated zonal flows (ZF) is the key problem to understanding the turbulence.  This has been called the ``drift wave-zonal flow'' problem \citet{diamond-zonal}.  To clarify the key nonlinear physics, a comparison is often drawn between ITG and electron temperature gradient (ETG) turbulence, which are linearly isomorphic: whereas ETG turbulence displays anisotropic streamers \cite{jenko-dorland-pop}, aligned radially with the temperature gradient, the ITG turbulence spectrum is characteristically smeared out among radial Fourier modes, resulting in a much more isotropic energy spectrum (and lower relative transport) -- the reason for this difference is ZFs.

It can be argued that the difference between ITG and ETG turbulence is even more fundamental: while an ETG mode is thought to saturate by energy transfer directly into secondary modes \cite{jenko-dorland-pop}, the energetic cost imposed on an ITG mode in the generation and sustainment of ZFs is small, and the real impact of the ZF is in its conservative ``shearing'' of the ITG mode; see \cite{plunk-njp}.  The term ``regulator'' is thus applied to ZFs, evoking the analogy of a special device (\eg a valve) that limits the operation of a machine.

Significant progress has been made recently in the understanding of ITG turbulence.  A cascade model was proposed by \citet{barnes-cb-itg} that eschews details of ITG mode saturation and ZF physics in favor of a simple set of scaling conjectures.  Although excellent agreement with numerical simulations was found, the theoretical explanation seems incongruous with both previous \cite{waltz-holland} and subsequent \cite{nakata} work, both of which support the paradigm of ZF regulation.  In particular, it was demonstrated \citet{nakata} that nonlinear energy transfer between ITG modes in the saturated steady state is dominated by wavenumber triads that are zonal-flow mediated (see Figure 11(b) of \cite{nakata}); in other words, the energy transfer between ITG modes is accounted for mostly by {\em ZF shearing}.  Adding further detail to the picture of ZF regulation, it was subsequently discovered \citet{plunk-njp} that the amplitude of ZFs in the turbulent state is controlled by energetic properties of the linear eigenmodes that drive the turbulence.  In particular, it was shown that the generation of ZFs by the inverse cascade of electrostatic energy (\ie its flow to large scales via nonlinear interaction) can be weakened or even reversed if the unstable modes possess sufficiently large free energy, a condition that is approached, \eg, at large temperature gradients (see Sec.~\ref{sat-sec}).

Despite the existence of a large body of research, the challenge remains to understand in a detailed and quantitative manner the processes by which the ITG turbulent state forms.   The present work addresses this challenge, focussing on the strongly nonlinear regime, well above marginal stability, where ZFs saturate by nonlinear decay (rather than by collisions or other linear damping mechanisms).  We propose a phenomenological cascade model, based on the conservative nonlinear transfer of energy to small scales, where it is ultimately dissipated.  The model is developed and tested as follows.  By performing electrostatic gyrokinetic simulations (employing the GENE code) with parameters similar to the Cyclone Base Case (CBC), we demonstrate (1) that the rate of energy injection in ITG turbulence is determined by the linear growth rate and (2) a properly defined ZF shearing rate balances with this rate at {\em energy-containing scales}.  We thus argue that saturation occurs by the nonlinear generation of ZFs and concurrent shearing of ITG modes.  The key new ingredient in this picture is that the ITG mode intensity needed to produce sufficiently strong ZFs is sensitive to the energy content of the ITG modes; we explain this physically using secondary and tertiary instability theory.  We propose a quantitative model that incorporates these ingredients and show that it agrees with the observed heat flux in the strongly-driven limit.

\section{Equations and definitions}We solve the electrostatic gyrokinetic equation for ions, which can be written (ignoring collisions) as

\begin{equation}
\frac{\partial \hat{h}}{\partial t} + \vpar \frac{\partial \hat{h}}{\partial z} + i \odt \hat{h}  = \sum_{\bf k'} \epsilon({\bf k}_{\perp}, {\bf k}_{\perp}^{\prime}) J_0\hat{\phi}({\bf k}_{\perp}') \hat{h}({\bf k}_{\perp}-{\bf k}_{\perp}') + (\frac{\partial}{\partial t} + i\ost)\frac{q\hat{\phi}}{T_0} J_0 f_M,\label{gk-eqn}
\end{equation}

\noindent where the coupling coefficient is $\epsilon({\bf k}_{\perp}, {\bf k}_{\perp}^{\prime}) = \Bo^{-1}(\hat{\bf b}\times{\bf k}_{\perp}^{\prime})\cdot{\bf k}_{\perp}$, with ${\bf B} = \hat{\bf b}\Bo = \bnabla\psi\times\bnabla\alpha$, where $\psi$ is the flux surface label, and $\alpha$ is the field line label.  The ion gyrocenter distribution is $\hat{h}({\bf k}_{\perp}, z, \varepsilon, \mu, t)$, with $z$ denoting the coordinate along the magnetic field line, ${\bf k}_{\perp} = k_{\psi}\bnabla\psi + k_{\alpha}\bnabla \alpha $ the wavenumber perpendicular to the magnetic field, an $\mu$ and $\varepsilon$ the magnetic moment and energy respectively.  The frequencies are defined $\odt = {\bf k}_{\perp}\cdot \hat{\bf b}\times[(\vperp^2/2)\bnabla \Bo + \vpar^2\hat{\bf b}\cdot\bnabla\hat{\bf b}]/\Omega_c$, and $\ost = \os[1 + \eta(v^2/\vT^2 - 3/2)]$, with $\eta = d\ln T_0/d\ln n_0$, $\os = (T_0k_{\alpha}/q)d\ln n_0/d\psi$, $\vpar = \hat{\bf b}\cdot{\bf v}$, $\vperp = |\hat{\bf b}\times{\bf v}|$, $J_0 = J_0(\kperp\vperp/\Omega_c)$ and $\Omega_c = q\Bo/m$, and $q = Ze$ the ion charge in terms of the absolute electron charge $e$.  The equilibrium ion distribution function is $f_M = n_0/(\vT^3\pi)^{3/2}\exp(-v^2/\vT^2)$, $\vT = \sqrt{2T_0/m}$, where $n_0$ and $T_0$ are the bulk ion density and temperature.  For the electron species we assume a modified Boltzmann response, so the quasi-neutrality constraint determining $\phi= \sum_{{\bf k}_{\perp}} \hat{\phi} \exp(i {\bf k}_{\perp}\cdot{\bf r})$ is

\begin{equation}
n_0\frac{q(\phi+\tau\nzo{\phi})}{T_0} = \int d^3{\bf v} \sum_{{\bf k}_{\perp}} \hat{h} J_0 \exp(i {\bf k}_{\perp}\cdot{\bf r}),
\end{equation}

\noindent where $\tau = T_0/(ZT_{e0})$, and $T_{e0}$ is the electron temperature.  The non-zonal part of the potential $\nzo{\phi}$ is defined

\begin{equation}
\nzo{\phi} = \phi - \zon{\phi},\label{nzo-def}
\end{equation}

\noindent and the zonal part $\zon{\phi}$ is defined by the usual flux surface average

\begin{eqnarray}\label{flux-avg-eqn}
\eqalign{\zon{\phi} &= \frac{\int_0^{2\pi} d\alpha \int dz \phi/B}{2\pi \int dz/B} \cr
&= \sum_{k_{\psi}}\exp(i k_\psi \psi)\frac{\int dz \hat{\phi}(k_\psi,k_\alpha = 0, z)/B}{\int dz/B} \cr
&= \sum_{k_{\psi}}\exp(i k_\psi \psi)\left< \hat{\phi}(k_\psi, k_\alpha = 0, z) \right>_{\parallel}}
\end{eqnarray}

\noindent In the final line we have introduced the field line average $\left<.\right>_{\parallel}$.  We will refer to $\zon{\phi}$ and $\nzo{\phi}$ respectively as the zonal and non-zonal parts of $\phi$.  In our theoretical arguments, we invoke the strongly ballooning limit (see \ref{strong-appx}), whereby the dynamics are nearly two-dimensional (\ie do not vary significantly along ${\bf B}$).  In this case we may use the local approximation $\odt \approx \od [\vpar^2/\vT^2 + \vperp^2/(2\vT^2)]$, and the local notation $\os = k_y\rho \vT/(\sqrt{2}L_n)$, where $\od = k_y\rho \sqrt{2}\vT/R$, $L_n^{-1} = d\ln n_0/dx$, $L_T^{-1} = d\ln T_0/dx$, $\eta = L_n/L_T$, $k_y = k_\alpha B|_{z = 0} (dx/d\psi)/\sqrt{\pi}$, $\rho = \vT/(\sqrt{2}\Omega|_{z = 0})$, and $R$ is the radius of curvature for the magnetic field.  Note that henceforth we will find it convenient to use ($k_x$, $k_y$) in place of ($k_\alpha$, $k_\psi$) to conceal the presence of magnetic geometry and generally simplify the presentation.

Numerical simulations are done with flux tube domains, which are three-dimensional (non-uniform in $z$).  These simulations are characterized by dimensionless parameters $\kappa = R/L_T$, $\tau$, $\eta$, and $\Lpar/R$, where we have introduced a single scale $\Lpar$ \footnote{For a generic equilibrium magnetic field, arbitrarily many quantities may be introduced to parameterize its spatial variation; however, by limiting ourselves to the CBC model geometry, it is sufficient to use the single parameter $\Lpar = \pi q R$, where $q$ is the safety factor.} to stand as the characteristic length scale associated with the variation of equilibrium quantities in the direction of the mean magnetic field.  Transport fluxes can depend on these dimensionless parameters but must obey gyro-Bohm scaling \cite{hagan-frieman}.

To define a turbulent cascade, we require an energy quantity, which, following convention, we take to be the free energy 

\begin{equation}
W = \int \frac{d^3{\bf r}}{V} \left[\int d^3{\bf v}\frac{T_0 h^2}{2 f_M} -  \frac{q^2n_0(\phi^2+\tau\nzo{\phi}^2)}{2T_0}\right],
\end{equation}

\noindent where $V$ is the system volume.  This quantity is conserved by all terms in the collisionless gyrokinetic equation (\ref{gk-eqn}) except the source term proportional to $\ost$. 

\section{Turbulence phenomenology}
\label{phenom-sec}

In order to describe ITG turbulence in terms of an energy cascade, we will need to use a ``phenomenological'' description, by which we mean a shorthand notation for describing the cascade process, \ie the injection and nonlinear turnover of energy, {\it etc}.  Although it may be useful to use a loosely-defined phenomenology to develop initial understanding, it is clearly preferable to use precisely defined quantities, as we will do here.  This sharpens the understanding and makes the theory more directly testable.  For fluid turbulence, it was suggested in Chapter 6 of \cite{frisch} that the velocity field at scale $\ell$, denoted $v_{\ell}$, should be defined as the root-mean-square-value of the field filtered around the wavenumber $\ell^{-1}$.  That is, writing ${\bf v} = \sum_{\bf k}\hat{\bf v}\exp(i{\bf k}\cdot{\bf r})$, and denoting the ensemble average (informally a time average) as $\left <.\right>$, we can relate $v_{\ell}$ to the RMS Fourier component as follows:

\begin{eqnarray}\label{phenom-def-eqn}
\eqalign{v_{\ell}  &\sim \sqrt{\left<\left|\displaystyle\sum_{k\;\sim\;1/\ell}\hat{\bf v}\exp(i{\bf k}\cdot{\bf r})\right|^2\right>} \cr
&= \sqrt{\displaystyle\sum_{k\;\sim\;1/\ell}\left<\left|\hat{\bf v}\right|^2\right>}\cr
&\sim \left(\frac{1}{\Delta k}\right)^{D/2} \sqrt{\int\limits_{k\;\sim\;1/\ell} dk k^{D-1} \left<\left|\hat{\bf v}\right|^2\right>}\cr
&\sim \left[\left(k/\Delta k\right)^{D/2} \hat{v}_{\mathrm{RMS}}\right]_{k = 1/\ell}}
\end{eqnarray}

\noindent where $\hat{v}_{\mathrm{RMS}} \equiv \sqrt{\left<\left|\hat{\bf v}\right|^2\right>}$, $D$ is the dimension of the space in which the cascade occurs, and $k \sim \ell^{-1}$ means $\ell^{-1} \leq |{\bf k}| < 2\ell^{-1}$.  Going between the first and second line we have assumed the turbulence is statistically homogeneous in space (Wiener--Khinchin theorem).  Note that the first line can be considered as a definition of $v_{\ell}$, but since this is a phenomenological quantity, it is only defined up to an overall multiplicative constant of order unity.  Thus the factor of 2 in the range of $k$ is somewhat arbitrary.  In deriving the final line we have assumed that the spectrum is such (\eg a power law) that the effect of integrating over $k$ simply introduces a factor of $k$.  The final line demonstrates that the number of modes that participate in local energy cascade is effectively $(k/\Delta k)^{D/2}$.

For present purposes, the electrostatic potential $\phi = \sum_{\bf k}\hat{\phi}\exp(i{\bf k}\cdot{\bf r})$ is the turbulent field, and we will need to divide it into zonal and non-zonal contributions, \ie $\phi = \nzo{\phi} + \zon{\phi}$ where recall $\zon{\phi}$ is the flux-surface average of $\phi$.  ITG turbulence is not isotropic, but, if one excludes the zonal component, it does have a uniform distribution in $k_x$ for $k_x \lesssim k_y$.  Thus we define $\nzo{\phi}_{\ell}$, the non-zonal electrostatic potential at scale $\ell$, as

\begin{equation}
\nzo{\phi}_{\ell} \equiv \left[\frac{k_y}{\sqrt{\Dkx\Dky}}\hat{\phi}_{\mathrm{RMS}}(k_x, k_y)\right]_{k_y = 1/\ell,\;\; k_x \leq k_y},\label{nzo-phi-def}
\end{equation}

\noindent where we have averaged over $z$, and excluded the zonal component, \ie%we have summed over $\kpar$, the wavenumber conjugate to $z$, \ie 

\begin{equation}
\hat{\phi}_{\mathrm{RMS}}(k_x, k_y) = \sqrt{\left<\left<\left|\hat{\phi}(k_x, k_y, z) - \delta_{k_y}\left<\hat{\phi}(k_x, k_y, z)\right>_{\parallel}\right|^2\right>_{\parallel}\right>},\label{phi-hat-2D-eqn}
\end{equation}

\noindent where $\delta_{k_y}$ is the discrete delta function.  Next, we define the zonal potential at scale $\ell$ as 

\begin{equation}
\zon{\phi}_{\ell} \equiv \left[\left(\frac{k_x}{\Dkx}\right)^{1/2} \phiHz_{\mathrm{RMS}}(k_x) \right]_{k_x = 1/\ell},\label{zon-phi-def}
\end{equation}

\noindent where

\begin{equation}
\phiHz_{\mathrm{RMS}}(k_x) = \sqrt{\left<\left|\left<\hat{\phi}(k_x, 0, z)\right>_{\parallel}\right|^2\right>}.\label{phiZ-def}
\end{equation}

\noindent Note that the dimensional factor $(k_x/\Dkx)^{1/2}$ reflects the fact that shearing by ZFs is a one-dimensional process, \ie the flow of energy is only in the $k_x$ direction.

If we now assume that the energy of turbulence peaks at a characteristic scale $\ello$, then the total energy density obtained by integrating the spectrum can be expressed in terms of the fields at that scale.  That is, following \Eref{phenom-def-eqn}, we can write $E = V^{-1}\int d{\bf r} |{\bf v}|^2/2 \sim v_{\ello}^2$, where $V$ denotes the system volume.  In a similar manner, the heat flux is expressed phenomenologically as

\begin{equation}
Q = \sum_{\bf k} \frac{n_0k_y}{B} \Im\left<\hat{\phi}^*\hat{\delta T}\right> \sim \frac{n_0 \phi_{\ello}\delta T_{\ello}}{B\ello}
\end{equation}

\noindent To obtain the final expression, we have assumed either that (a) the phase of $\hat{\phi}$ and $\hat{\delta T}$ are uncorrelated, or (b) that $\hat{\phi}$ and $\hat{\delta T}$ are not systematically in-phase to the degree that a significant cancellation occurs.

\section{Energy injection and nonlinear transfer}
\begin{figure*}
\includegraphics[width=0.95\textwidth]{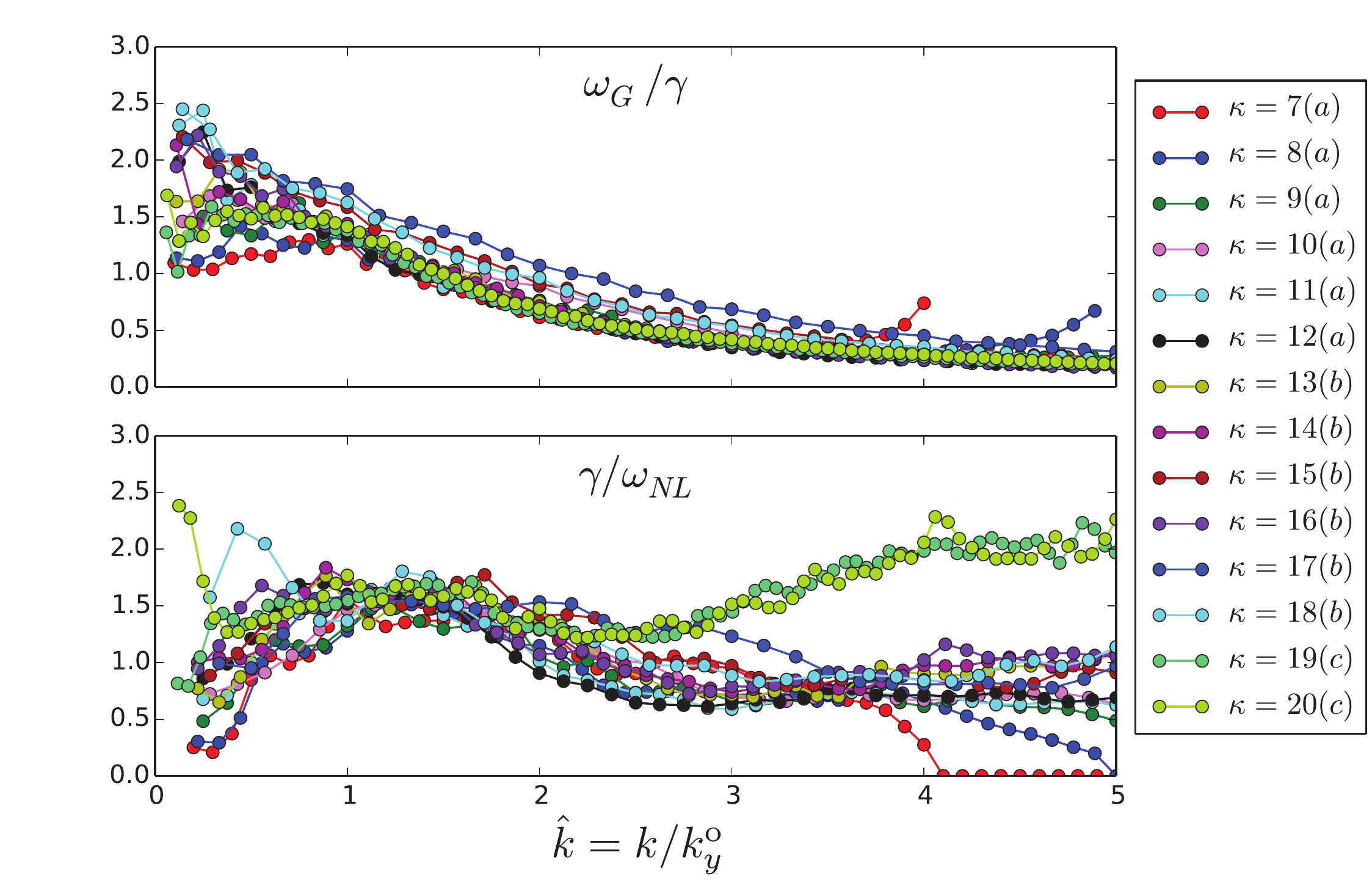}
\caption{Comparison of linear and nonlinear rates associated with energy flows.  The simulations are grouped, \ie $7(a)$-$12(a)$, $13(b)$-$18(b)$, and $19(c)$-$20(c)$, to signify a doubling of the $x$-$y$ domain size from one group to the next; this increase is required due to the decrease of $\kyo$ with increasing $\kappa$ \cite{barnes-cb-itg}.}
\label{freq-compare-fig}
\end{figure*}

Our model assumes that the turbulence has two key properties: (P1) the nonlinear turnover rate is determined by a local (in wavenumber space) ZF-mediated transfer of free energy and (P2) the injection of energy is largely determined by the linear instability.\footnote{It has been noted by several authors that gyrokinetic turbulence remains ``nearly linear,'' as reflected by signatures of the linear eigenmodes in the fully saturated state of nonlinear simulations \cite{dannert-jenko, goerler-jenko, hatch-terry}.}  As we develop the model below, these assumptions will be made more precise, and their validity will be tested using numerical simulations; see \Fref{freq-compare-fig}.

Because large scales dominate the transport spectrum of ITG turbulence, let us consider the long-wavelength fluid limit of gyrokinetics, $\kperp^2\rho^2 \ll 1$.  At these scales, nonlinear phase-mixing \cite{dorland-hammett-93} and other finite-Larmor-radius (FLR) effects are weak, and the density moment of the gyrokinetic equation (\ref{gk-eqn}) yields (see \cite{plunk-njp}, Eqn.~D.6)

\begin{equation}
\frac{\partial\nzo{\phi}}{\partial t} + \frac{1}{\Bo}(\hat{\bf b}\times\bnabla\zon{\phi}) \cdot \bnabla \nzo{\phi} = \mathcal{S}.\label{nz-fluid-eqn}
\end{equation}

\noindent Here we recall that $\nzo{\phi}$ is the non-zonal part of $\phi$ (see \Eref{nzo-def}), and $\mathcal{S}$ represents linear source terms; the equation for the evolution of the zonal part $\zon{\phi}$ is not shown here.  The form of \Eref{nz-fluid-eqn} reflects the nonlocal response of the electrons, which travel rapidly along the magnetic field.  It is well-known that this modification to the traditional Boltzmann electron response causes strong ZF generation.  As \Eref{nz-fluid-eqn} reveals, another consequence of the electron response is that, for $\kperp^2\rho^2 \ll 1$, the nonlinear evolution of the non-zonal potential is due entirely to conservative ``shearing'' by ZFs.  Note however that the linear coupling ($\mathcal{S}$) to moments other than density (\eg parallel ion flow and perpendicular ion temperature fluctuations) underlies the mechanism of linear instability, and so one cannot know {\it a priori} whether the nonlinear interactions involving the non-zonal component of such moments might lead to mode saturation instead of the ZF shearing mechanism.  Evidence from the literature helps rule out the possibility that these so-called ``drift-wave/drift-wave'' (DW/DW) interactions might be important:  First, when ITG or ETG modes saturate without ZFs, the process generates fine structure parallel to the magnetic field, and a strongly anisotropic turbulence spectrum perpendicular to the magnetic field, as a consequence of the Cowley secondary instability \cite{jenko-dorland-pop, cowley-kulsrud}.  This, however, is not typical in ITG turbulence with CBC-like parameters, on which we focus here; an exception is found at low $\tau$ as discussed later.  Second, and on a more empirical note, these DW/DW interactions have been investigated in numerical simulations and found not to contribute significantly to nonlinear transfer \cite{waltz-holland, nakata}, as previously discussed.

In short, we argue that \Eref{nz-fluid-eqn} captures the dominant nonlinear energy transfer mechanism.  This mechanism can be thought of as a one-dimensional cascades in $k_x$-space, \ie energy is transported in $k_x$-space under the action of zonal shearing, but not transferred significantly in $k_y$-space (P1).

Let us describe this process phenomenologically.  We first assume {\em locality}, \ie that the nonlinear interactions tend to involve wavenumbers of a similar magnitude.  Locality in gyrokinetic turbulence is well-established \cite{tatsuno-jpf, banon-prl, teaca-prl}.  The turbulence can then be described in terms of the quantities $\nzo{\phi}_{\ell}$ and $\zon{\phi}_{\ell}$, as discussed in \Sref{phenom-sec}, bearing in mind their relationship to the RMS potential $|\hat{\phi}(k_x, k_y, z)|_{\mathrm{RMS}}$, henceforth abbreviated as simply $\hat{\phi}$.

Now we let us proceed to defining the nonlinear turnover rate $\omegaNL$.  \Eref{nz-fluid-eqn} describes a one-dimensional shearing process.  We model this as an energy cascade, where the nonlinear turnover (at the rate $\omegaNL$) is due to the zonal $E \times B$ velocity acting on non-zonal fluctuations.  Using \Eref{zon-phi-def} we can obtain the nonlinear turnover rate at perpendicular wavenumber ($k_x$, $k_y$), \ie

\begin{equation}
\frac{\partial }{\partial t}\Biggr |_{\mathrm{NL}} \hat{\phi}(k_x, k_y) \sim \omegaNL \hat{\phi}(k_x, k_y)
\end{equation}

\noindent where

\begin{equation}
\omegaNL \sim \frac{k_x^{\mathrm{z}} k_y}{\Bo}\sqrt{\frac{k_x^{\mathrm{z}}}{\Dkx}}\phiHz(k_x^{\mathrm{z}}),\label{oNL-def-eqn}
\end{equation}

\noindent where $\phiHz$ is defined in \Eref{phiZ-def}, $\Dkx$ is the wavenumber spacing, $k_x^{\mathrm{z}}$ is the wavenumber of the ZF, and $(k_x, k_y)$ correspond to the non-zonal mode that is being sheared.  The meaning of \Eref{oNL-def-eqn} is that the energy in mode $(k_x, k_y)$ turns over at the rate $\omegaNL$ due to shearing by ZFs having wavenumbers around $k_x^{\mathrm{z}}$.  However, since the most unstable ITG modes satisfy $k_x \lesssim k_y$, the further assumption of {\em locality} ($k_x^{\mathrm{z}} \sim k_y$) is sufficient to express the required nonlinear turnover rate of such modes in terms of the single wavenumber $k \sim k_x^{\mathrm{z}} \sim k_x \sim k_y$:

\begin{equation}
\omegaNL \sim \frac{k^{5/2}}{\Bo\sqrt{\Dkx}}\phiHz(k).\label{oNL-eqn}
\end{equation}

\noindent To compare with this rate, we define the linear growth at wavenumber $k$ as

\begin{equation}
\gammaL(k) \equiv \gammaL(k_y = k, k_x = 0).\label{gamma-def-eqn}
\end{equation}

Let us test how well the assumptions of the above model are satisfied by numerical simulations of the turbulence; \Fref{freq-compare-fig} summarizes our findings.  For these plots the wavenumber is normalized to the peak of the heat transport spectrum (approximately the same as the peak of the free energy spectrum), denoted $\kyo$ or alternatively $\ello \equiv 1/\kyo$.  Note that if the nonlinear turnover of the turbulence increases sufficiently fast in $k_y$, then the wavenumber $\kyo$ will be set by the low-$k_y$ cutoff \cite{jenko-varenna} of the (dominant) instability, rather than its peak; indeed $\kyo$ is sometimes determined this way in quasilinear transport estimates \cite{citrin}. The bottom panel of \Fref{freq-compare-fig} compares the maximum linear growth rate for ITG modes (\ie for $\bkperp  = (0, k)$) with the nonlinear turnover rate $\omegaNL(k)$.  The top panel compares the same growth rate with the rate of injection of free energy in the fully developed turbulence, $\omega_G$.  This quantity is formally defined by splitting the energy budget equation into linear and nonlinear contributions at each $k_y$ (see \citep{banon-pop} for details): $dW(k_y)/dt = (dW(k_y)/dt)_\mathrm{L} + (dW(k_y)/dt)_\mathrm{NL}$, where $W(k_y) = \sum_{k_x, k_z} W(k_x, k_y, k_z)$, and subscripts `L' and `NL' denote contributions from linear and nonlinear terms respectively.  Then $\omega_G$ is

\begin{equation}
\oG(k_y) \equiv \frac{1}{W(k_y)} \left(\frac{dW(k_y)}{dt}\right)_{\mathrm{L}}.\label{oG-eqn}
\end{equation}

\noindent The frequency $\omega_G$ is the correct quantity to compare with the nonlinear frequency $\omegaNL$ since nonlinear turnover refers explicitly to the transfer of energy in $k$-space.  Our propositions (P1) and (P2) can now be succinctly stated as 

\begin{equation}
\omegaNL \sim \gammaL \sim \omega _G,\label{omega-balance-eqn}
\end{equation}

\noindent with values given by \Eref{oNL-eqn}, \Eref{gamma-def-eqn} and \Eref{oG-eqn}.  However, these are not equalities, so their validity is tested not by how close the frequency ratios are to unity, but by whether or not they systematically change with the variation of parameters.  \Fref{freq-compare-fig} demonstrates that the balance $\omegaNL \sim \gammaL$ is satisfied uniformly well over a significant range of $\kappa$.  However, there is a significant variation in $k$; this may reflect a different cascade process at small scales, as we will discuss later.  The balance $\omega_G \sim \gammaL$, is also satisfied uniformly well over $\kappa$.\footnote{In fact, in the purely linear system $\omega_G = 2\gammaL$ would be satisfied exactly (after a transient period) because the growth in energy would be due to the most unstable modes at each ${\bf k}$.  Thus, the small value of the ratio $\omega_G/\gammaL$ at high $k$ signals nonlinear behavior, and may be due the nonlinear excitation of the damped spectrum of modes \cite{hatch-terry}.}

\section{Saturation of the turbulence}\label{sat-sec}Having validated the assumptions of our model, let us now complete our description of the saturation process.  Using \Eref{oNL-eqn} and \Eref{gamma-def-eqn}, the balance $\gamma \sim \omegaNL$ implies a saturation rule for the zonal potential:

\begin{equation}
\phiHz(k) \sim \gammaL(k) \frac{\Bo}{k^2}\sqrt{\frac{\Dkx}{k}},\label{sat-rule-eqn}
\end{equation}

\noindent where $\phiHz(k)$ was defined in \Eref{phiZ-def}.  Note that this saturation rule differs from that of \citet{waltz-holland} in the mode-counting factor $(k_x/\Dkx)^{1/2}$, and differs from that of \cite{hahm-hammett} in that it does not assume timescale separation.

Equipped with \Eref{sat-rule-eqn}, we must now determine the relative amplitude of non-zonal fluctuations to complete the description of the saturated state.  One might naively guess that the nonlinear rate associated with DW energy turnover, $\omegaNL \sim \ell^{-2}\zon{\phi}_{\ell}/\Bo$, should balance with the nonlinear rate of the zonal potential, $\omegaNLz \sim \ell^{-2}\nzo{\phi}^2_{\ell}/(\zon{\phi}_{\ell}\Bo)$ (this follows from the vorticity equation for $\zon{\phi}$, \ie the zonal part of Eqn.~(D.18) of \cite{plunk-njp}), which would lead to the relative saturation amplitude $\nzo{\phi}_{\ell} \sim \zon{\phi}_{\ell}$.  This turns out to be incorrect: it is missing a constant of proportionality, \ie we should instead write $\nzo{\phi}_{\ell} \sim \alpha \zon{\phi}_{\ell}$, or in $k$-space (see \Eref{nzo-phi-def} and \Eref{zon-phi-def}) we have

\begin{equation}
\hat{\phi}(0, k) \sim \alpha \phiHz(k)\sqrt{\Dky/k}.\label{zonal-resp-factor}
\end{equation}

\noindent \Eref{sat-rule-eqn} and \Eref{zonal-resp-factor} raise a subtle conceptual issue:  A zonal Fourier component only scales with the system size in the $x$ dimension, $\Dkx$.  This is in contrast to a Fourier component in isotropic turbulence, which must scale with the full system volume, \ie also with $\Dky$ and $\Dkz$, because the number of modes per unit volume in $k$-space increases with $\Delta k$ and so each individual Fourier mode becomes less important.  Consequently, for isotropic turbulence, in the limit that any one of the dimensions of the system is large, any collection of Fourier components lying on a surface in $k$-space must occupy a vanishing fraction of the $k$-space volume and could be neglected without influencing the system dynamics.  ZFs, however, occupy a single line in $k$-space, and therefore if ZFs are to act as the regulator of the turbulence they must always be important and cannot diminish with increasing $\Dky$ or $\Dkz$.  For this reason, one should not directly compare zonal and non-zonal Fourier modes as their relative amplitudes depends on system size, as specified in \Eref{zonal-resp-factor}.

Returning to the meaning of \Eref{zonal-resp-factor}, the factor $\alpha$ quantifies a certain ``non-universality'': in addition the scale ($1/k$) and rate ($\gammaL$), it turns out that the saturation process also depends on an additional energetic property of the ITG mode itself, which causes the nonlinear response of the ZFs to weaken as $\kappa$ increases.\footnote{Note that as $\kappa$ increases, the ZFs weaken while the turbulence simultaneously goes to larger scales, causing the nonlinear terms due to the FLR (neglected in deriving \Eref{nz-fluid-eqn}) to also weaken.  Thus, zonal shearing can remain the dominant nonlinear transfer mechanism.}

This effect was described by \citet{plunk-njp} as a cascade reversal, controlled by an energy ratio parameter $W/E$, where for our purposes $E \approx \tau \Ephi$ where $\Ephi = n_0\sum_{{\bf k}_{\perp}}q^2|\hat{\phi}|^2/(2T_0)$.  To obtain a quantitative estimate for $\alpha$, we will take a slightly different approach via the primary/secondary/tertiary instability framework conceived by \citet{cowley-kulsrud, rogers-prl}.  This approach allows one to probe the nonlinear physics by linearizing the dynamics about a nonlinearly-motivated initial condition.  The secondary instability is calculated by linearizing the gyrokinetic equation about a large-amplitude ``primary'' ITG mode, and the tertiary mode is calculated by doing the same about a large-amplitude zonal mode.  We argue that the relative amplitude of zonal and non-zonal fluctuations is set by balancing secondary and tertiary growth rates, \ie $\gamma_s$ and $\gamma_t$.  The ratio of zonal and non-zonal amplitudes is set according to this balance, because if it deviates significantly, then one instability will overwhelm the other and guide the system back into a state of balance.  

Although the secondary instability is usually expressed in terms of an isolated Fourier amplitude, we use the quantity $\nzo{\phi}_{\ell}$ to reflect the turbulent state.  Thus, we take 

\begin{equation}
\gamma_s \sim k^{\mathrm{p}}_y \rho k^{\mathrm{s}}_x\vT \left[\frac{q\nzo{\phi}_{\ell}}{T_0}\right]_{\ell = 1/k^{\mathrm{s}}_x},
\end{equation}

\noindent where $k^{\mathrm{s}}_x$ is the wavenumber of the secondary mode and $k^{\mathrm{p}}_y$ is the wavenumber of the primary mode; see Appendix D.5 of \cite{plunk-njp}.  Defining $\delta T_{\perp} = n_0^{-1} \int d^3{\bf v} \; m \vperp^2(h - q\phi/T)$ we take 

\begin{equation}
\gamma_t \sim k^{\mathrm{t}}_y\rho \tau^{-1/2} \vT (k^{\mathrm{s}}_x)^2\rho \left[\frac{(q\zon{\phi}_{\ell}\zon{\delta T}_{\perp\ell})^{1/2}}{T_0}\right]_{\ell = 1/k^{\mathrm{s}}_x},
\end{equation}

\noindent where $k^{\mathrm{t}}_y$ is the wavenumber of the tertiary \cite{rogers-prl}.\footnote{Note that our calculation neglects the effects that stabilize the tertiary at low-$\kappa$ (such as finite $\kpar$), so its validity is limited to values of $\kappa$ significantly larger than the nonlinear critical gradient.  However it seems generalizable to the weakly supercritical regime if more complete tertiary mode physics is included.}   Now balancing these expressions for $\gamma_s \sim \gamma_t$, and using $k^{\mathrm{s}}_x\rho \sim k^{\mathrm{p}}_y\rho \sim k\rho$ (\ie locality), and \Eref{zonal-resp-factor}, we find

\begin{equation}
\alpha \sim \sqrt{\frac{k^{\mathrm{t}}_y\rho\zon{\delta T}_{\perp}}{q\zon{\phi}\tau}}.
\end{equation}

\begin{figure}
\includegraphics[width=0.95\columnwidth]{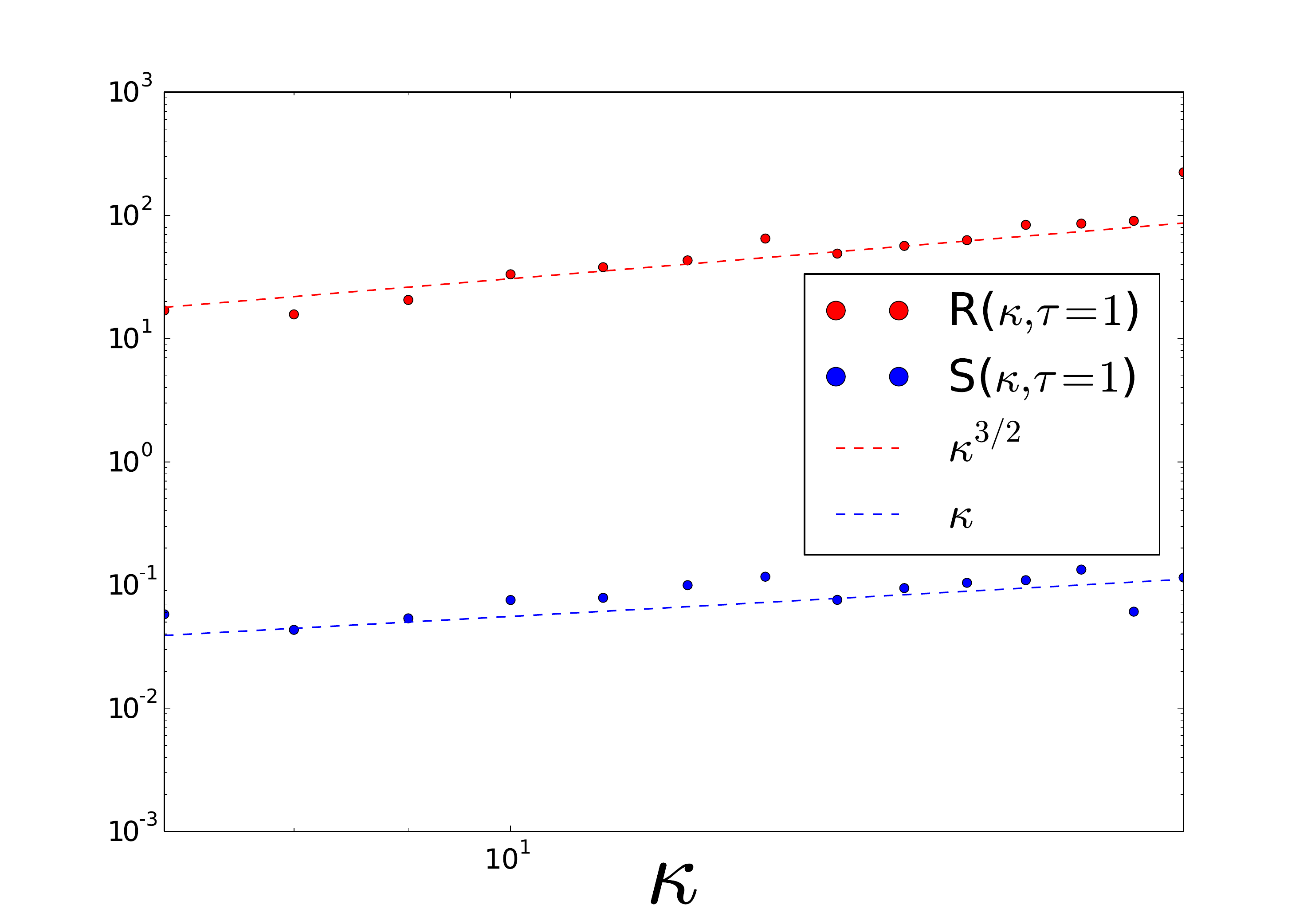}
\caption{Nonlinear response of ZFs.  Here $R = \Dky^{-1}\langle\sum_{k_x}|\hat{\phi}(k_x, \kyo)|^2\rangle/\langle\sum_{k_x}|\phiHz(k_x)|^2\rangle$, where $\langle .\rangle$ is an average over time.  Also plotted is $S = \langle\sum_{k_x}|\THz(k_x)|^2\rangle/\langle q^2\sum_{k_x}|\phiHz(k_x)|^2\rangle$, where $\THz$ is defined analogously to \Eref{phiZ-def}.}
\label{zon-nzo-fig}
\end{figure}

\noindent This expression depends on an unknown, the wavenumber $k^{\mathrm{t}}_y$ of the tertiary.  It was found by \cite{rogers-prl} that $k^{\mathrm{t}}_y \sim \sqrt{k^{\mathrm{s}}_x/\rho}$ gives a maximal growth rate.  However, there is a large range of scales accessible to the tertiary mode and the mode of peak growth rate is not necessarily the most important, so it is not clear how $k^{\mathrm{t}}_y$ should be chosen.\footnote{Consider also that the nonlinear transfer of electrostatic energy shown in Figure 2(b) of \cite{banon-pop} is nonlocal and involves a broad range of scales.  This hints that the nonlinear zonal decay process may involve a range of scales.}  Furthermore, the fully gyrokinetic calculation made by \cite{plunk-njp} demonstrated that $\gamma_t$ and $k^{\mathrm{t}}_y$ both depend on the ratio of zonal temperature to zonal density.  We assume, then, that the quantity $k^{\mathrm{t}}_y\rho$ can be expressed as a power of $\zon{\delta T}_{\perp}/q\zon{\phi}$, and find $k^{\mathrm{t}}_y\rho \sim \zon{\delta T}_{\perp}/q\zon{\phi}$ agrees with simulation results (see \Fref{zon-nzo-fig}).  Thus we obtain

\begin{equation}
\alpha \sim \frac{1}{\sqrt{\tau}}\frac{\zon{\delta T}_{\perp}}{q\zon{\phi}}.\label{alpha-eqn}
\end{equation}

\section{Strongly driven limit} Now we can consider the strongly driven (large $\kappa$) limit, for which we can evaluate several quantities according to linear theory:

\begin{eqnarray}
\gammaL = \sqrt{\osT\od/\tau} \sim k_y\rho \sqrt{\kappa/\tau}(\vT/R),\label{gamma-tor-eqn}\\
\nzo{\delta T}/q\nzo{\phi} \sim \nzo{\delta T}_{\perp}/q\nzo{\phi} \sim \sqrt{\kappa \tau},\label{T-phi-ratio-eqn}\\
\kyo \sim \rho^{-1}\kappa^{-1/2}R/\Lpar.\label{outerscale-eqn}
\end{eqnarray}

The estimates \ref{gamma-tor-eqn} and \ref{T-phi-ratio-eqn} are determined by the strongly ballooning toroidal ITG mode (see \ie \cite{plunk-itg} and \ref{strong-appx}).  The expression \ref{outerscale-eqn} can be obtained by balancing the sound wave transit rate with the growth rate \ref{gamma-tor-eqn}, \ie $c_s/\Lpar = \vT\tau^{-1/2}/\Lpar\sim \gammaL(\kyo)$; note that this balance is similar to what \cite{barnes-cb-itg} call ``critical balance,'' and both yield the proportionality $\kyo \propto R/\Lpar$, though they disagree on the scaling in $\kappa$.  The wavenumber $\kyo$ corresponds to the transition from the toroidal branch to the subdominant slab branch, which is clearly observed in linear simulations.  Naively, one might expect the linear estimate for $\delta T/q\phi$, \ie \Eref{T-phi-ratio-eqn}, to be satisfied for the zonal component.  However, although the $\kappa$-dependence is satisfied, the $\tau$-dependence is not, at least for the range of parameters that we have investigated; instead we observe $\zon{\delta T}_{\perp}/q\zon{\phi} \sim \sqrt{\kappa}$, which using \Eref{alpha-eqn} yields

\begin{equation}
\alpha \sim \sqrt{\kappa/\tau}.\label{alpha-large-kappa-eqn}
\end{equation}

\noindent The scalings above are supported by \Fref{zon-nzo-fig}.  For the plotted quantity $R = \Dky^{-1}\langle\sum_{k_x}|\hat{\phi}(k_x, \kyo)|^2\rangle/\langle\sum_{k_x}|\phiHz(k_x)|^2\rangle$, we expect from \Eref{zonal-resp-factor}, \Eref{outerscale-eqn}, and \Eref{alpha-large-kappa-eqn} that $R \propto \alpha^2/\kyo \propto \kappa^{3/2}$.  For $S = \langle\sum_{k_x}|\THz(k_x)|^2\rangle/\langle q^2\sum_{k_x}|\phiHz(k_x)|^2\rangle$, \Eref{T-phi-ratio-eqn} implies $S \propto \kappa$.

\begin{figure}
\includegraphics[width=0.95\columnwidth]{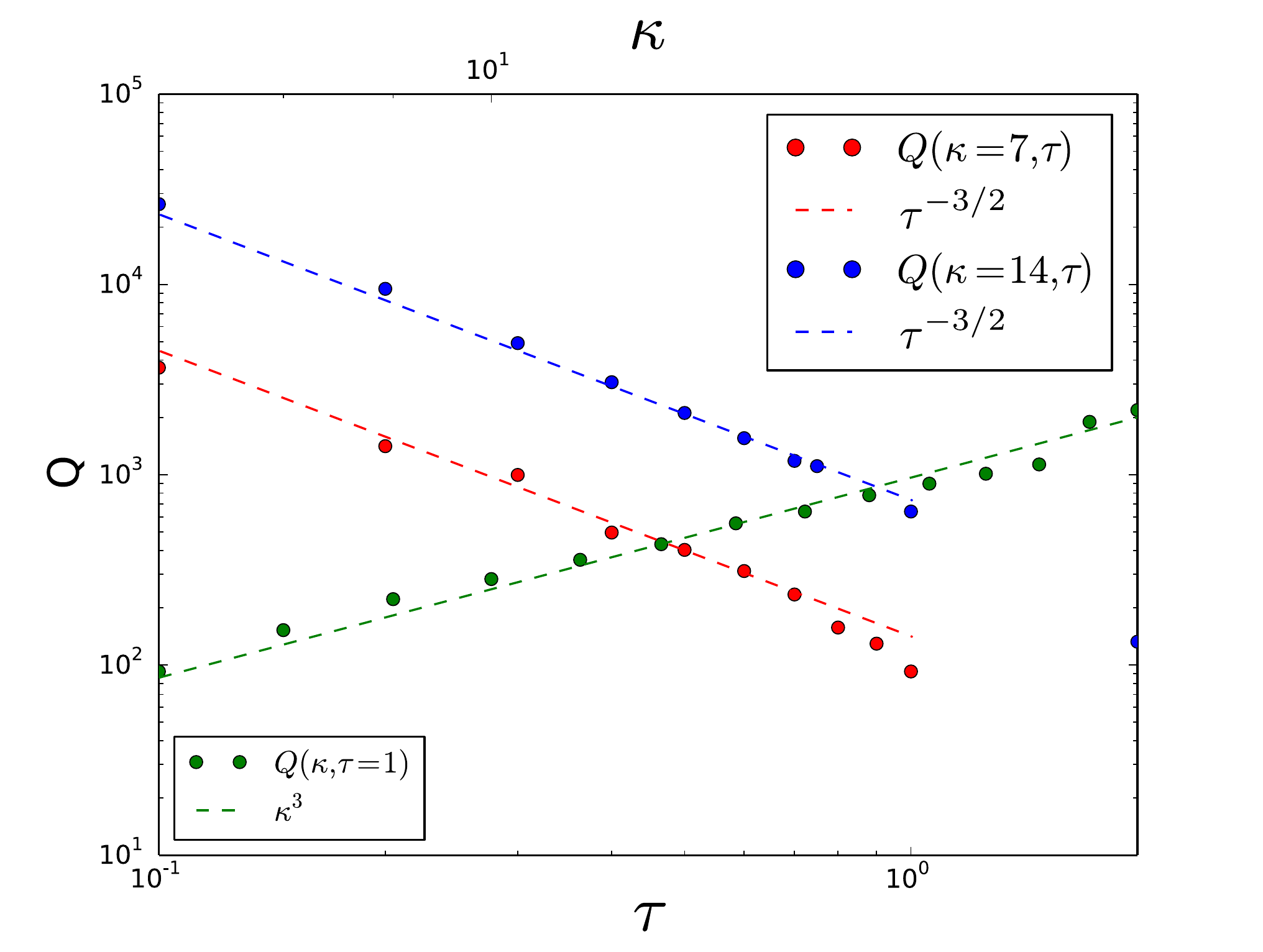}
\caption{Scaling of heat flux with dimensionless parameters.  Dashed lines correspond to theoretical predictions for the strongly driven limit.  Note that the top horizontal axis is used for $\kappa$ (green data points) and the bottom horizontal axis is used for $\tau$ (blue and red data points).}
\label{Q-scaling-fig}
\end{figure}

Now, using Equations \ref{sat-rule-eqn}, \ref{zonal-resp-factor}, \ref{gamma-tor-eqn}, \ref{T-phi-ratio-eqn}, \ref{outerscale-eqn}, and \ref{alpha-large-kappa-eqn}, the heat flux may be calculated as $Q \sim n_0 \nzo{\phi}_{\ello}\nzo{\delta T}_{\ello}/(B\ello)$ (see \Sref{phenom-sec}), yielding

\begin{equation}
Q \sim n_0 T_0\vT\rho^2\tau^{-3/2}\left(\frac{\Lpar}{L_T^3}\right),\label{q-scaling-eqn}
\end{equation}

\noindent which differs from \cite{barnes-cb-itg} in its $\tau$ dependence; this difference is significant because it reflects completely different saturation physics than that assumed by \cite{barnes-cb-itg}.  Our simulations obey the scaling in \Eref{q-scaling-eqn}, as shown in \Fref{Q-scaling-fig}.  Note that although nothing fundamental prohibits applying this to $\tau > 1$, the ITG mode becomes stabilized for the values of $\kappa$ that we used.  Note also that deviation from the theoretical line is expected at sufficiently low-$\tau$ because, given fixed $\kyo$, the weakening of the ZFs must ultimately cause nonlinear interactions among the ITG modes (DWs) to become important.  Indeed, although all our other simulations show the signature of dominant zonal shearing, \ie a flat energy distribution for $k_x \leq k_y$, the spectrum of the low-$\tau$ outliers exhibit peaking at low-$k_x$ and associated streamer-like structures.

\begin{figure}
\includegraphics[width=0.95\columnwidth]{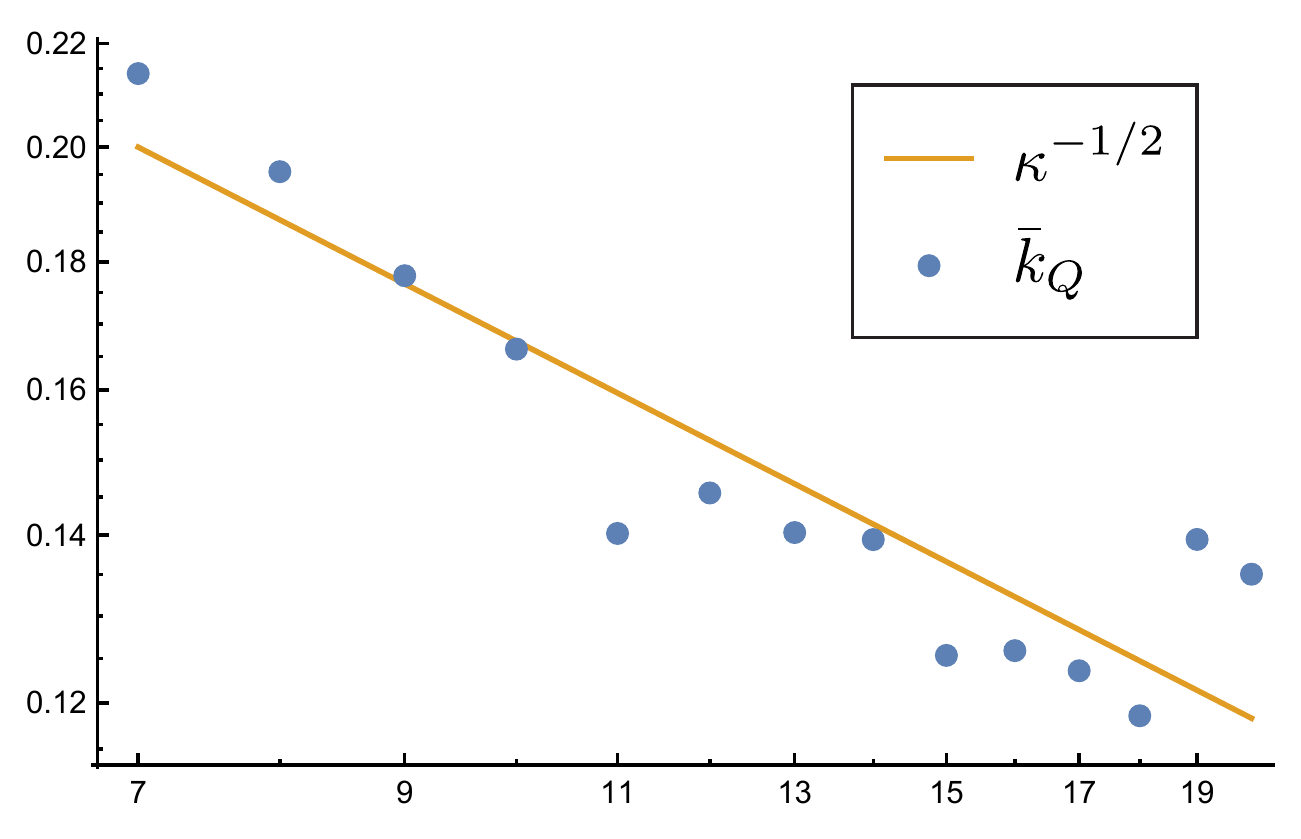}
\caption{Measure of outer scale wavenumber $\kyo$.}
\label{ratio-outer-scales-fig}
\end{figure}

An important ingredient in the calculation of the overall saturation and heat flux is the wavenumber $\kyo$, which corresponds to the outer scale of the turbulence.  As a measure for $\kyo$, the quantity $\bar{k}_Q = \sum_{k_y} Q(k_y)/\sum_{k_y} Q(k_y)k_y^{-1}$ is plotted in \Fref{ratio-outer-scales-fig} and compared with the theoretical prediction $\kappa^{-1/2}$ of \Eref{outerscale-eqn}.  This prediction differs from the outer-scale estimate of \citep{barnes-cb-itg}, that find $\bar{k}_{\phi} \propto \kappa^{-1}$, where $\bar{k}_{\phi} = \sum_{k_y} \Ephi(k_y)/\sum_{k_y} \Ephi(k_y)k_y^{-1}$.  We suspect that these two definitions give different results because the spectrum of electrostatic energy $\Ephi$ tends to a constant at low $k_y$ and so the factor of $k_y^{-1}$ preferentially weights large scales  in the sum $\sum_{k_y} \Ephi(k_y)k_y^{-1}$, \ie the lowest available wavenumber dominates the computation of $\bar{k}_{\phi}$.  Thus we conclude that $\bar{k}_{Q}$ is the correct measure of the dominant wavenumber contributing to heat flux.

\section{Spectra}

It is noteworthy that detailed knowledge of the energy spectrum is not important for determining the scaling of bulk properties of the turbulence ($W$, $Q$, \etc) with respect to the system parameters.  This is because the integral of a power-law spectrum will only introduce an overall dimensionless factor, corresponding to the steepness of the spectrum.  Nevertheless, it is interesting that some universal behavior is exhibited, at least in the regime considered here.

Formally, we define the spectrum of zonal and non-zonal energies as $\zEphi(k_x) =  \Dkx^{-1}|\phiHz(k_x)|^2$ and $\nEphi(k_x, k_y) =  \Dkx^{-1}\Dky^{-1}|\hat{\phi}(k_x, k_y)|^2$, respectively.   The spectrum of zonal energy is predicted immediately by the saturation rule \Eref{sat-rule-eqn}, and the degree to which this is confirmed is established by the second panel of \Fref{freq-compare-fig}.  The dependence $\gammaL \propto k_y$ then implies that $\zEphi \propto k_x^{-3}$ for at least some range.

The non-zonal spectrum at $k_y > \kyo$ depends on the saturation process at those wavenumbers.  As we have noted, the balance between injection, nonlinear turnover by zonal shearing, and linear growth is not uniformly satisfied in $k$.  Furthermore, there is evidence that zonal shearing only provides a small part of energy transfer at inertial-range (small) scales \citep{teaca-2014-pop}.  It is thus possible that an isotropic cascade (in $\bkperp$) overtakes the zonal-mediated transfer at these scales.  However we argue that an isotropic cascade is not possible around $\kyo$ because the associated nonlinear turnover rate is too weak to account for the observed energy injection; if, for example, this rate was computed according to the FLR corrections of \Eref{nz-fluid-eqn}, one would obtain $\omega_{\mathrm{NL}}^{\mathrm{iso}} \equiv \rho^2\ell^{-4}\nzo{\phi}_{\ell}/B \ll \gammaL$ for $\ell \sim 1/\kyo$.  The spectral scaling $\int dk_x \nEphi \equiv \nEphi(k_y) \propto k_y^{-7/3}$ \cite{barnes-cb-itg} seems consistent with our numerical results in the large $\kappa$ limit, although low-$\kappa$ clearly has a steeper spectrum.

\section{Discussion}We have presented a model of the saturation process of ITG turbulence, which agrees with numerical simulations in a detailed and quantitative way.  We have made this description as explicit and physically transparent as possible, and a key part of our theoretical argument is a set of linearly calculable (primary, secondary, and tertiary) instabilities.  We hope that these features will make it possible to generalize our results in the future.

Based on our findings, the physical picture of ITG turbulence is as follows.  The saturation of the turbulent state is determined by a cascade process at an outer scale.  The injection of free energy into the turbulence is determined by the linear growth rate of the ITG mode, and the flux of this energy is carried to high-$k_x$ by the shearing action of ZFs, which we have described as a one-dimensional cascade.  The amplitude of the DWs relative to the ZFs is sensitive to the amount of free energy (relative to the electrostatic energy) that is entering the cascade, which is a feature of the ITG mode itself.  This energy ratio is a fundamentally kinetic quantity, as it measures the amount of excitation of velocity-space structure in the distribution function.  This property of the ITG mode can be quantified via the ratio of the temperature and potential fluctuations, and combined with the saturation rules, and outer scale estimate, yields the observed scaling of the ion heat flux $Q$.

Generally, our findings support the paradigm of ZF regulation, and we have shown that it applies even in the strongly-driven regime.  Interestingly, we have seen that the ZFs are capable of regulating the turbulence, even as their amplitude (relative to DWs) is diminished.  This fact should serve as a warning against using the relative amplitude of the ZFs as a measure of their strength.  Instead the dynamical importance of ZFs is properly evaluated by comparing the rate of zonal shearing to the rate of energy injection by linear instability.

One issue that merits further investigation is the nonlinear decay process of the ZFs.  We have cited the tertiary instability of \cite{rogers-prl} and given an argument for how it balances with ZF generation process (secondary instability), but a detailed description that accounts for a range of interacting scales, especially near marginal stability of the ITG mode, remains an open problem.

As a final point, we should note that our work is clearly only valid when the nonlinear decay of ZFs dominates over collisional decay.  This is the regime of strong turbulence, corresponding to sufficiently large temperature gradient, \ie those above the nonlinear critical gradient of Dimits \cite{dimits}.  It is not clear what regime is most relevant for present experiments, but future fusion devices with lower collisionality will exhibit lower transport in the weakly driven regime (in which ZF decay is collisional) and so the strong regime should be more accessible.  Furthermore, if collisional decay and nonlinear decay must both be included to model a given experiment, it should prove useful to have a good understanding of the fully nonlinear regime.

\section{Acknowledgements}We gratefully acknowledge Per Helander for helpful comments, and the Wolfgang Pauli Institute for hosting a series of workshops on Gyrokinetics.  The research leading to these results has received funding from the European Research Council under the European UnionÕs Seventh Framework Programme (FP7/2007-2013)/ERC Grant Agreement No. 277870.

\appendix

\section{Properties of the strongly-driven toroidal ITG mode}\label{strong-appx}

The term ``strongly ballooning'' describes an asymptotic limit of the full three-dimensional ballooning mode calculation, whereby the two conditions are simultaneously satisfied: (1) the mode is strongly localized at the outboard midplane (location of bad curvature) and (2) the transit frequency (\ie the rate associated with ion streaming a distance equal to the scale of the mode parallel to the magnetic field) is small.  Because these two conditions may appear contradictory, it is useful to show explicitly that they can be satisfied simultaneously.  A demonstration of this has been given in a recent work \cite{plunk-itg}.

The strongly ballooning limit is essentially the strongly driven limit, which is also a non-resonant limit.  The following local dispersion relation for the toroidal branch of the ITG mode is valid ($J_0 = 1$):

\begin{equation}
(1 + \tau)\frac{q\phi}{T_0} = \frac{1}{n_0}\int d^2{\bf v} h,
\end{equation}

\noindent where $\int d^2{\bf v} = 2\pi \int \vperp d\vperp \vpar$.  The distribution $h$ is

\begin{equation}
h = \frac{\omega - \ost}{\omega - \odt}\frac{q\phi}{T_0}f_0,
\end{equation}

\noindent where $\ost = \ost + \osT[v^2/\vT^2 - 3/2]$ and $\odt = \od [\vpar^2/\vT^2 + \vperp^2/(2\vT^2)]$.  The strongly driven limit can be explicitly written as $\odt \ll \omega \ll \osT$.  Taking also $\os \ll \omega$, the solution can be written $\omega \approx i\sqrt{\osT\od/\tau}$.  To calculate the perturbed pressure (as it appears in the heat transport flux) we need only retain the dominant contribution, \ie $h = -(\osT[v^2/\vT^2 - 3/2]/\omega)(q\phi/T_0)f_0$, so that

\begin{eqnarray}
\delta T &= \frac{1}{n_0} \int d^2{\bf v}\frac{mv^2}{2} h\\
& \sim \frac{\osT}{\omega} q\phi \sim i q\phi \sqrt{\kappa\tau} 
\end{eqnarray}

\noindent Note that the temperature is $\pi/2$ out of phase with the potential, which means it contributes fully to transport.

\section*{References}
\bibliographystyle{unsrt}
\bibliography{zf-cascade-PPCF-v3}

\begin{thebibliography}{10}

\bibitem{diamond-zonal}
P~H Diamond, S-I Itoh, K~Itoh, and T~S Hahm.
\newblock Zonal flows in plasma—a review.
\newblock {\em Plasma Phys. Control. Fusion}, 47(5):R35, 2005.

\bibitem{jenko-dorland-pop}
F.~Jenko, W.~Dorland, M.~Kotschenreuther, and B.~N. Rogers.
\newblock Electron temperature gradient driven turbulence.
\newblock {\em Physics of Plasmas (1994-present)}, 7(5):1904--1910, 2000.

\bibitem{plunk-njp}
G~G Plunk, T~Tatsuno, and W~Dorland.
\newblock Considering fluctuation energy as a measure of gyrokinetic
  turbulence.
\newblock {\em New J. Phys.}, 14(10):103030, 2012.

\bibitem{barnes-cb-itg}
M.~Barnes, F.~I. Parra, and A.~A. Schekochihin.
\newblock Critically balanced ion temperature gradient turbulence in fusion
  plasmas.
\newblock {\em Phys. Rev. Lett.}, 107:115003, Sep 2011.

\bibitem{waltz-holland}
R.~E. Waltz and C.~Holland.
\newblock Numerical experiments on the drift wave--zonal flow paradigm for
  nonlinear saturation.
\newblock {\em Phys. Plasmas}, 15(12):122503, 2008.

\bibitem{nakata}
M.~Nakata, T.-H. Watanabe, and H.~Sugama.
\newblock Nonlinear entropy transfer via zonal flows in gyrokinetic plasma
  turbulence.
\newblock {\em Phys. Plasmas}, 19(2):022303, 2012.

\bibitem{hagan-frieman}
W.~K. Hagan and E.~A. Frieman.
\newblock Nonlinear gyrokinetic theory, the direct interaction approximation,
  and anomalous thermal transport in tokamaks.
\newblock {\em Phys. Fluids}, 29(11):3635--3638, 1986.

\bibitem{frisch}
U.~Frisch.
\newblock {\em Turbulence: The Legacy of A. N. Kolmogorov}.
\newblock Cambridge University Press, 1995.

\bibitem{dannert-jenko}
Tilman Dannert and Frank Jenko.
\newblock Gyrokinetic simulation of collisionless trapped-electron mode
  turbulence.
\newblock {\em Phys. Plasmas}, 12(7):--, 2005.

\bibitem{goerler-jenko}
T.~G{\"o}rler and F.~Jenko.
\newblock Multiscale features of density and frequency spectra from nonlinear
  gyrokinetics.
\newblock {\em Phys. Plasmas}, 15(10):--, 2008.

\bibitem{hatch-terry}
D.~R. Hatch, P.~W. Terry, F.~Jenko, F.~Merz, and W.~M. Nevins.
\newblock Saturation of gyrokinetic turbulence through damped eigenmodes.
\newblock {\em Phys. Rev. Lett.}, 106:115003, Mar 2011.

\bibitem{dorland-hammett-93}
W.~Dorland and G.~W. Hammett.
\newblock Gyrofluid turbulence models with kinetic effects.
\newblock {\em Phys. Fluids B}, 5(3):812--835, 1993.

\bibitem{cowley-kulsrud}
S.~C. Cowley, R.~M. Kulsrud, and R.~Sudan.
\newblock Considerations of ion-temperature-gradient-driven turbulence.
\newblock {\em Phys. Fluids B}, 3(10):2767, 1991.

\bibitem{tatsuno-jpf}
T.~Tatsuno, M.~Barnes, S.~C. Cowley, W.~Dorland, G.~G. Howes, R.~Numata, G.~G.
  Plunk, and A.~A. Schekochihin.
\newblock Gyrokinetic simulation of entropy cascade in two-dimensional
  electrostatic turbulence.
\newblock {\em J. Plasma Fusion Res. Ser.}, 9:509, 2010.

\bibitem{banon-prl}
A.~Ba{\~n}{\'o}n~Navarro, P.~Morel, M.~Albrecht-Marc, D.~Carati, F.~Merz,
  T.~G\"orler, and F.~Jenko.
\newblock Free energy cascade in gyrokinetic turbulence.
\newblock {\em Phys. Rev. Lett.}, 106:055001, Jan 2011.

\bibitem{teaca-prl}
Bogdan Teaca, Alejandro~Ba{\~n}{\'o}n Navarro, Frank Jenko, Stephan Brunner,
  and Laurent Villard.
\newblock Locality and universality in gyrokinetic turbulence.
\newblock {\em Phys. Rev. Lett.}, 109:235003, Dec 2012.

\bibitem{jenko-varenna}
F.~Jenko, W.~Dorland, B.~Scott, and D.~Strintzi.
\newblock Simulation and theory of temperature gradient driven turbulence.
\newblock In O.~Sauter J.W.~Connor and E.~Sindoni, editors, {\em Theory of
  Fusion Plasmas}, page 157. Societa Italiana di Fisica, Bologna, 2002.

\bibitem{citrin}
J.~Citrin, C.~Bourdelle, P.~Cottier, D.~F. Escande, \"{O}.~D. G\"{u}rcan, D.~R.
  Hatch, G.~M.~D. Hogeweij, F.~Jenko, and M.~J. Pueschel.
\newblock Quasilinear transport modelling at low magnetic shear.
\newblock {\em Physics of Plasmas}, 19(6):062305, 2012.

\bibitem{banon-pop}
A.~Ba{\~n}\'on Navarro, P.~Morel, M.~Albrecht-Marc, D.~Carati, F.~Merz,
  T.~G\"orler, and F.~Jenko.
\newblock Free energy balance in gyrokinetic turbulence.
\newblock {\em Phys. Plasmas}, 18(9):092303, 2011.

\bibitem{hahm-hammett}
T.~S. Hahm, M.~A. Beer, Z.~Lin, G.~W. Hammett, W.~W. Lee, and W.~M. Tang.
\newblock Shearing rate of time-dependent e x b flow.
\newblock {\em Physics of Plasmas}, 6(3):922--926, 1999.

\bibitem{rogers-prl}
B.~N. Rogers, W.~Dorland, and M.~Kotschenreuther.
\newblock Generation and stability of zonal flows in ion-temperature-gradient
  mode turbulence.
\newblock {\em Phys. Rev. Lett.}, 85(25):5336--5339, Dec 2000.

\bibitem{plunk-itg}
G.~G. Plunk, P.~Helander, P.~Xanthopoulos, and J.~W. Connor.
\newblock Collisionless microinstabilities in stellarators. iii. the
  ion-temperature-gradient mode.
\newblock {\em Phys. Plasmas}, 21(3):--, 2014.

\bibitem{teaca-2014-pop}
Bogdan Teaca, Alejandro~Ba{\~n}{\'o}n Navarro, and Frank Jenko.
\newblock The energetic coupling of scales in gyrokinetic plasma turbulence.
\newblock {\em Phys. Plasmas}, 21(7):--, 2014.

\bibitem{dimits}
A.~M. Dimits, G.~Bateman, M.~A. Beer, B.~I. Cohen, W.~Dorland, G.~W. Hammett,
  C.~Kim, J.~E. Kinsey, M.~Kotschenreuther, A.~H. Kritz, L.~L. Lao,
  J.~Mandrekas, W.~M. Nevins, S.~E. Parker, A.~J. Redd, D.~E. Shumaker,
  R.~Sydora, and J.~Weiland.
\newblock Comparisons and physics basis of tokamak transport models and
  turbulence simulations.
\newblock {\em Phys. Plasmas}, 7(3):969, 2000.

\end{thebibliography}

\end{document}